\documentstyle[twocolumn,psfig,floats,aps]{revtex}

\begin{document}
\twocolumn[\hsize\textwidth\columnwidth\hsize\csname
@twocolumnfalse\endcsname
\title{Coexistence of charge density waves, 
bond order waves and spin density waves 
in quasi-one dimensional charge transfer salts}
\author{J.~Riera$^{a,b}$ and D.~Poilblanc$^{a}$}
\address{
$^a$Instituto de F\'{\i}sica Rosario, Consejo Nacional de
Investigaciones 
Cient\'{\i}ficas y T\'ecnicas, y Departamento de F\'{\i}sica,
Universidad Nacional de Rosario, Avenida Pellegrini 250, 2000-Rosario,
Argentina\\
$^b$Laboratoire de Physique Quantique \& UMR--CNRS 5626,
Universit\'e Paul Sabatier, F-31062 Toulouse, France
}
\date{\today}
\maketitle

\begin{abstract}
Charge, spin, as well as lattice instabilities 
are investigated in 
isolated or weakly coupled chains of correlated electrons 
at quarter-filling.
Our analysis is based on extended Hubbard models
including nearest neighbor repulsion and 
Peierls coupling to lattice degrees of freedom.
While treating the electronic quantum fluctuations exactly,
the lattice structure is optimized self-consistently. 
We show that, generically, isolated chains undergo instabilities
towards coexisting charge density waves (CDW) 
and bond order waves (BOW) insulating spin-gapped
phases. The spin and charge gaps of the BOW-CDW phase are computed.
In the presence of an interchain
magnetic coupling spin density waves phases including a CDW or a BOW 
component are also found.
Our results are discussed in the context
of insulating charge transfer salts.

\end{abstract}

\vskip2pc]



Quasi-one dimensional correlated electrons systems at 
quarter-filling
show fascinating physical properties. 
A widely studied class of 
such materials are the so-called organic
charge transfer salts like the Bechgaard salts (TMTSF)$_2$X
or their sulfur analogs (TMTTF)$_2$X 
(X=PF$_6$, AsF$_6$)~\cite{review,review2,review3}.
These systems which consist of stacks of organics molecules forming
weakly coupled one dimensional (1D) chains 
exhibit, at low temperature, a large variety of exotic phases 
such as triplet superconducting~\cite{triplet}, 
spin density wave (SDW), charge density wave (CDW)
and spin-Peierls (SP) phases~\cite{ESR}. 
The compounds of the sulfur (TMTTF)$_2$X series show 
strong charge localization at rather high temperature 
$T_\rho$ (as signalled e.g. by transport measurements)
as well as a weak (to moderate) dimerization along the 
stacks~\cite{review,review2}.
The insulating behavior observed in this regime has been interpreted
as 1D CDW fluctuations~\cite{review3} or using
a Hubbard chain Hamiltonian with an explicit dimerization~\cite{Penc}.
At a significantly lower temperature $T_{SP}$ (typically
$T_{SP}\sim 15 K$) a SP transition occurs together 
with a tetramerization along the chains~\cite{review,ESR}.
In this system the interplay between charge ordering and lattice
instability is poorly understood.
In particular whether the tetramerization (connected to the spin gap) is 
occurring simultaneously with a CDW
transition is still controversial. So far the existing theory of the SP 
phase~\cite{Dumoulin} do not consider the possibility of a coexisting
$2k_F$ CDW. 
Interestingly enough, CDW fluctuations were seen 
by X-ray diffuse scattering~\cite{Pouget_96} in the SDW phase of 
the (more metallic) (TMTSF)$_2$PF$_6$ compound. 

In this paper, our aim is to investigate by numerical
exact diagonalization (ED) techniques, the interplay between 
the electron repulsion and the electron-phonon
coupling in the case of an adiabatic lattice.
We focus on the competition or the cooperative behavior between 
charge ordering and lattice instabilities.
Such a problem has been addressed in a number of previous 
studies~\cite{Mazumdar1,Mazumdar2} where several
interesting modulated phases have been proposed.
However, so far, no systematic investigation of the full phase diagrams
has been carried out. Indeed, the suggested translation symmetry 
broken states~\cite{Mazumdar2} were found 
in a restricted variational space. We shall here re-examine these issues
in order to determine the absolute stability of the various competing
phases.

We use extended
1D Hubbard models at quarter-filling (${\bar n}=1/2$)
coupled with some (classical) phonon field,
\begin{eqnarray}
H_{1D}&=& \sum_{i,\sigma} t(i) (c_{i;\sigma}^\dagger
c_{i+1;\sigma}+h.c.)+U\sum_{i} n_{i;\uparrow}n_{i;\downarrow}
\nonumber \\
&+& V\sum_{i} n_{i}n_{i+1} + H_{\rm ph} \, ,
\label{Ham}
\end{eqnarray}
where $n_{i;\sigma}=c_{i;\sigma}^\dagger c_{i;\sigma}$ and 
$n_i=n_{i;\uparrow}+n_{i;\downarrow}$.
We have included a nearest neighbor (NN) interaction $V$ as its role
will become clear in the following.
Local deformations of the molecules 
can produce changes of the on-site (or molecular) orbitals 
energies and can simply be taken into account 
by a Holstein term $\sum_{i} n_{i}\delta_{i}$
while assuming a constant hopping integral $t(i)=t$
and an elastic energy cost $\frac{1}{2}K\sum_{i}\delta_{i}^2$. 
This effect has been studied numerically in a different 
context~\cite{NaVO} but it is relatively small in 
the case of the organic systems here considered.
In contrast to the above on-site deformation, the positions 
of the intercalated anions can couple strongly to the electrons
especially through modulations of the single particle hoppings
along the chains of the Peierls type,
\begin{equation}
t(i)=t\, (1+\delta^B_i)
\end{equation}
with an elastic energy 
\begin{equation}
H_{\rm ph}\equiv H_{\rm elas}=\frac{1}{2}K_B \sum_i (\delta^B_i)^2\, .
\end{equation}
The electron-lattice couplings have been absorbed in the 
re-definition of the displacements $\delta^B_i$
so that the strength of the lattice coupling scales like $1/K_B$
(also $t$ is set to 1).
We shall first consider the case of the {\it isolated}
Hubbard chain and, next, the role of 
an inter-chain magnetic coupling (in mean field).

In contrast to previous treatments~\cite{Mazumdar2}
our numerical method enable us to obtain the lowest energy
equilibrium lattice configuration without making any assumption
on the broken symmetry ground state (GS). In particular,
no super-cell order is imposed {\it a priori} and the GS 
configuration is obtained through a self-consistent procedure.
Indeed, the total energy functionals $E(\{\delta_{i}^B\})$ 
can be minimized with respect to the sets of distortions
$\{\delta_{i}^B\}$ by solving the following non-linear 
coupled equations,
\begin{eqnarray}
K_B\delta^B_{i} &+& t\big< c_{i;\sigma}^\dagger
c_{i+1;\sigma} + h.c.\big> =0\, .
\label{non_linear}
\end{eqnarray}
Here $\big<...\big>$ is the GS mean value obtained by ED (using
the Lanczos algorithm) of Hamiltonian (\ref{Ham}) on cyclic $L$-site
rings (with $L$ up to 16 sites). Since the second
term depends implicitly on the distortion pattern
$\{\delta_{i}^B\}$, 
Eq.~(\ref{non_linear}) can be solved by a
regular iterative procedure\cite{Dobry_Riera}.
A similar approach has been applied to the case of the on-site 
Holstein coupling~\cite{NaVO}.
We should stress here that within the intrinsic limitations of the
method (adiabatic lattice and finiteness of the system) our
resolution of the problem is basically exact (numerical accuracy
better than $10^{-7}$). Note also that once a small adiabatic
lattice coupling is included finite size effects become quite small. 


Before discussing our main results on the Hubbard-Peierls 
chain, let us briefly describe the phase diagram of
the Hubbard-Holstein chain~\cite{NaVO} in order to 
introduce the generic types of CDW states.
At quarter-filling, the Fermi wave vector is given by
$q_{2k_F}=\frac{\pi}{2}$ so that, at small $U$, one expects an
instability towards a 
$2k_F$ CDW state of wavevector $\lambda_{2k_F}=4a$ ($a$ is the
lattice spacing) mediated by the electron-lattice coupling. 
In contrast, for large U, the system becomes more similar to a
gas of interacting spinless fermions (SF) and 
the instability is likely to occur at wavevector $2k_F^{SF}=4k_F$.
More generally, we can parameterize the relative charge density
modulation as,
\begin{equation}
\frac{\Delta n_i}{\bar n}=\rho_{4k_F} \cos{(2\pi\frac{r_i}{2a})}
+\rho_{2k_F} \cos{(2\pi\frac{r_i}{4a}+\Phi_{2k_F})}
\, ,
\label{charge_density}
\end{equation}
where $\Delta n_i=\big<n_i\big>-{\bar n}$.
Complete phase diagrams of the Hubbard-Holstein chain
have been established in Ref.~\cite{NaVO} 
and we briefly summarize them here. For $V=0$
the metallic uniform U phase ($\rho_{2k_F}=\rho_{4k_F}=0$) is
restricted to a region at small lattice coupling.
Above a critical line $(1/K)_U$, three different insulating 
CDW phases can be distinguished; (i) at small $U$, a $2k_F$ CDW phase
($\rho_{4k_F}= 0$) centered on the sites, i.e. with $\Phi_{2k_F}=0$;
(ii) at intermediate $U$ (in the range 4--8), a 
{\it bond-centered} $2k_F$ CDW phase (i.e. with
$\Phi_{2k_F}=\frac{\pi}{4}$);
(iii) at large U, a $4k_F$ CDW ($\rho_{2k_F}= 0$).
As discussed in Ref.~\cite{NaVO}, a small NN repulsion
suppresses completely the intermediate phase and enlarges the
region of stability of the $4k_F$ CDW phase.
Although these CDW might have some relevance to the 
low temperature phase of the (TMTTF)$_2$X family, their
charge modulations should couple strongly to the anion potential. 
Therefore, we investigate next the role of a Peierls coupling. 


\begin{figure}
\vspace{-0.2truecm} 
\begin{center}
\psfig{figure=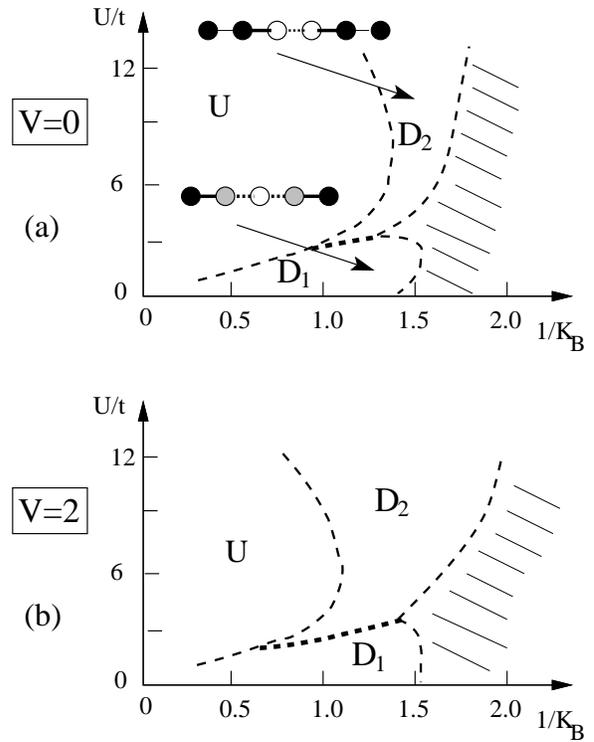,width=8truecm,angle=0}
\end{center}
\caption{
Typical ($U/t$,$1/K_B$) phase diagrams of a $\frac{1}{4}$-filled 
Hubbard-Peierls chain without (a) or with a nearest neighbor 
Coulomb repulsion $V$ (b) obtained from ED of small periodic chains.
Hashed regions are unphysical.
}
\label{Hubbard_Peierls}
\end{figure}

\begin{figure}[t]
\vspace{-0.2truecm} 
\begin{center}
\psfig{figure=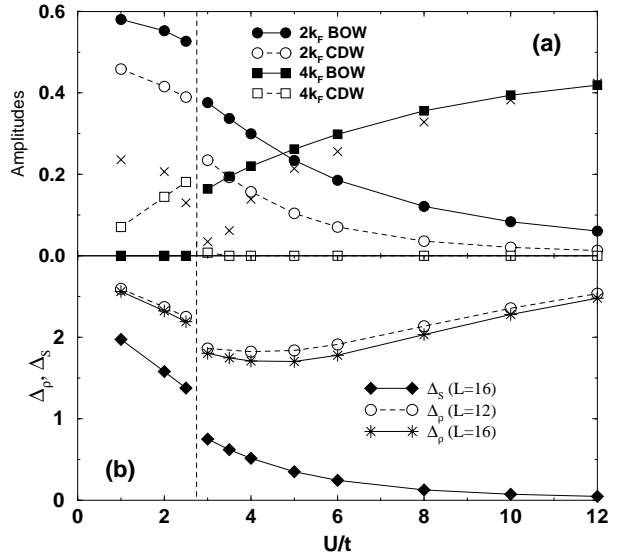,width=8truecm,angle=-90}
\end{center}
\caption{
(a) Amplitudes of the CDW ($\rho_{2k_F}$ and $\rho_{4k_F}$) and BOW 
($\delta_{2k_F}^B$ and $\delta_{4k_F}^B$) components versus $U/t$ for
$(K_B)^{-1}=1.25$ and $V=2$ (computed on a 16-site ring). 
Open (close) symbols correspond to CDW (BOW). Note that, although
identical symbols have been used, the 2k$_F$ orders are in fact
{\it different}, i.e. have different phases $\Phi_{2k_F}$ and 
$\Phi_{2k_F}^B$ in the $D_1$ and $D_2$ phases (see text).
The crosses indicate the energy difference (in absolute value)
between the states with these two patterns.
(b) Charge ($\Delta_\rho$) and spin ($\Delta_S$) gaps
in unit of $t$ vs $U/t$ computed on 12- and 16-site rings
($\Delta_S$ on $L=12$ and $L=16$ are undistinguishable).
}
\label{amplitudes}
\end{figure}

The phase diagrams as a function of the on-site repulsion $U/t$ and the 
Peierls coupling $1/K_B$ are shown in Figs.~\ref{Hubbard_Peierls}(a,b)
for $V=0$ and $V=2$. For too large electron-lattice coupling 
the linear coupling approximation breaks down and our model becomes
unphysical (hashed regions) so that we shall restrict 
to small and intermediate values of $1/K_B$. 
For intermediate values of $1/K_B$, the uniform state is unstable
towards translation symmetry broken states. Such
bond order wave (BOW) states are characterized by a modulation of 
the hopping amplitudes of the form, 
\begin{equation}
\delta_i^B=\delta^B_{4k_F} \cos{(2\pi\frac{r_i}{2a})}+
\delta^B_{2k_F} \cos{(2\pi\frac{r_i}{4a}+\Phi_{2k_F}^B)}
\, .
\end{equation}
Generically, we find that BOW coexist with weaker charge modulations
(CDW) given by (\ref{charge_density}).
The amplitudes of the bond and charge modulations are shown in  
Fig.~\ref{amplitudes}(a) for a fixed electron-lattice coupling $K_B=0.8$. 
Very small finite size effects are observed and calculations on 12- and
16-site rings give almost identical results.
Two different types of structures are stable; 
(i) in the weak coupling regime (let's say $U/t<3$), a strong 
2k$_F$ BOW with $\Phi_{2k_F}^B=\pi/4$ 
i.e. corresponding to a X--X--Y--Y type of sequence of the bonds
occurs.  This modulation coexists with a weaker 2k$_F$ site-centered 
CDW (A--B--${\bar{\rm A}}$--B type of sequence of the 
on-site charge densities, 
A and ${\bar{\rm A}}$ corresponding to opposite values of $\Delta n_i$) 
and an even weaker 4k$_F$ (i.e. A--B--A--B) CDW component ($D_1$ phase);
(ii) at larger $U/t$, the $D_2$ phase corresponds to the superposition of
a lattice dimerization (4k$_F$ BOW) together with a tetramerization
(2k$_F$ BOW with $\Phi_{2k_F}^B=0$). In other words, among the, let's say, 
weak bonds 
of the dimerized state, one every two becomes weaker (or stronger) so that 
electrons become weakly bound in singlet pairs on next NN bonds.
The $D_2$ phase is therefore a simple realization of the above mentioned
SP phase~\cite{Caron}.
Interestingly enough, we observe that the tetramerization leads to a weak
coexisting $2k_F$ CDW component correspond to a A--A--B--B charge
modulation (i.e. with $\Phi_{2k_F}=\pi/4$). 
Note also that the boundary between the $D_1$ and $D_2$ phases is 
a first-order transition line as it is clear from the discontinuity of the
various order parameters seen in Fig.~\ref{amplitudes}(a).
We have also included in this plot the energy difference between the
states corresponding to $D1$ and $D2$ patterns. This difference is in
general (except at large $U$) considerably smaller than charge and
spin excitations (see below).

To complete our study we have also computed the charge ($\Delta_\rho$)
and the spin ($\Delta_S$) gaps in the $D_1$ and $D_2$ phases 
for the same set of parameters as shown in Fig.~\ref{amplitudes}(b).
As for the Fourier amplitudes, finite size effects are almost negligible 
especially for $\Delta_S$. 
Clearly $\Delta_S$ follows closely the magnitude of the 2k$_F$ BOW-CDW.
For large-U (and large dimerization) 
the system behaves qualitatively like a spin-1/2 antiferromagnet (since
the electrons are localized 
on the strong bonds) and the spin gap is expected to vanish in this limit.
In contrast, in the $D_1$ phase, electrons are strongly localized in pairs
on 2 adjacent strong bonds (i.e. on 3 sites) so that $\Delta_S\sim t$.
The charge gap, also shown in Fig.~\ref{amplitudes},
has a minimum at intermediate $U$ in the region 
corresponding to the cross-over from dominant 2k$_F$ to
dominant 4k$_F$ BOW-CDW.
Note that both charge and spin gaps are discontinuous at the first
order transition between $D_1$ and $D_2$.

Our study shows that the electronic correlations (both $U$ and $V$) are 
essential to stabilize the $D_2$ phase which, we believe, is a fair 
realization of the SP phase of the (TMTTF)$_2$X material. 
Note that the 2k$_F$ (4k$_F$) Fourier components are suppressed 
(increased) as $U/t$ increases as seen in Fig.~\ref{amplitudes}(a) 
and in agreement with previous numerical calculations of the
on-site charge-density and NN charge-transfer response functions
of the extended Hubbard model~\cite{Hirsch}. 


Lastly, we investigate the role of an inter-chain coupling which 
is relevant e.g. in the case of (TMTTF)$_2$Br (or (TMTTF)$_2$PF$_6$
under pressure).
These systems which are less anisotropic than (TMTTF)$_2$PF$_6$
at ambient pressure exhibit an 
antiferromagnetic (AF) phase at low temperature. In an insulating
regime, due to the presence of a charge gap $\Delta_\rho$,
the inter-chain single particle hopping $t_\perp$ is believed to be
irrelevant~\cite{review3}. 
Therefore, we shall only consider a transverse magnetic
coupling $J_\perp$ (typically $J_\perp\sim t_\perp^2/\Delta_\rho$)
in mean field approximation. 
Our previous method can be straightforwardly extended to include this
inter-chain coupling by adding to (\ref{Ham}) an extra term as,
\begin{equation}
H_\perp=\frac{1}{2}\sum_i H_i (n_{i;\uparrow}-n_{i;\downarrow}),
\end{equation}
where the local fields $H_i$ are determined self-consistently 
(simultaneously with the $\delta_i^B$ bond modulations) from an
additional set of non-linear equations,
\begin{equation}
H_i=\frac{J_\perp}{2}\big<(n_{i;\uparrow}-n_{i;\downarrow})\big>\, .
\end{equation}

\begin{figure}
\begin{center}
\vspace{-0.2truecm} 
\psfig{figure=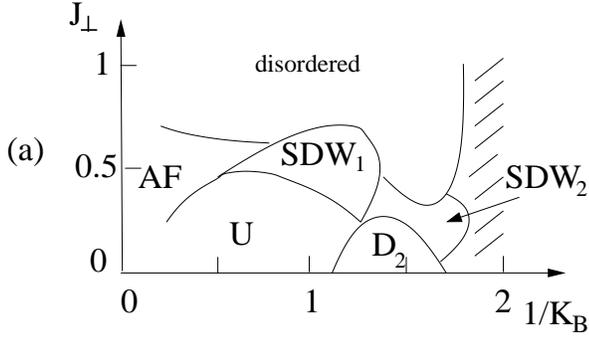,width=8truecm,angle=0}
\end{center}
\caption{
Phase diagram of weakly coupled $\frac{1}{4}$-filled 
Hubbard-Peierls chains as a function of 
$1/K_B$ and $J_\perp$ for $U$=6, $V=2$. 
Hashed regions are unphysical.
}
\label{AF}
\end{figure}

\begin{figure}
\begin{center}
\vspace{-0.2truecm} 
\hspace{0.4cm}\psfig{figure=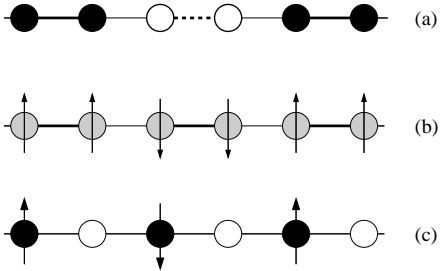,width=6truecm,angle=0}
\end{center}
\caption{
Pictures of the various phases of
antiferromagnetically coupled chains showing the bond modulation
(thick lines are strong bonds, dashed lines are weak bonds), the
charge modulation
(full and open circles correspond to excess and depression of charge
respectively) and the local spin densities (arrows).
(a) dimerized/tretramerized $D_2$ phase; (b) Spin density wave SDW$_1$
phase; (c) antiferromagnetic phase.
}
\label{sdw}
\end{figure}

Our results on coupled Hubbard-Peierls chains are summarized in
Fig.~\ref{AF}. Besides the $D_2$ phase (Fig.~\ref{sdw}(a))
which is stable at small $J_\perp$ new magnetic
phases depicted in Figs.~\ref{sdw}(b,c) appear; (i) at small
electron-lattice coupling an antiferromagnetic phase consisting of a 
site-centered 4k$_F$ CDW with a finite spin density on the
sites carrying an excess charge (see Fig.~\ref{sdw}(c)); (ii) at
larger values of $1/K_B$,
a superposition of a dimerization (4k$_F$ BOW) with a 2k$_F$ 
{\it bond-centered} SDW order~\cite{note_SDW} 
(see  Fig.~\ref{sdw}(b)). Note that only the spin densities of these
two magnetic phases could be obtained by adding two out-of-phase 
$2k_F$ CDW for the spins up and down so that the 
4k$_F$ CDW and BOW components should really be considered as extra 
coexisting orders.
In a small region of the parameter space, a more exotic 
magnetic SDW$_2$ phase (not shown) has been stabilized on our
clusters. This phase contains all three CDW, SDW and BOW
components with large supercells of the order of our chain lengths.
Finally, we note that the region of stability of the D$_2$ phase should
be extended by a small interchain bond coupling~\cite{Caron}.

To summarize, the role of Peierls electron-lattice
couplings has been
investigated in the adiabatic approximation in quarter-filled
isolated or weakly coupled one-dimensional Hubbard chains. A numerical
method based on ED techniques supplemented by a self-consistent
procedure has been used to determine the various phase diagrams
as a function of the strengths of the lattice coupling and the
Coulomb repulsion. We have shown that,
generically, lattice modulations (BOW) are always accompanied
by CDW's of weaker amplitudes. In addition, at intermediate and
large on-site Coulomb repulsion, the lattice modulation consists of a
superposition of a $4k_F$ (dimerization) and a $2k_F$ (tetramerization) 
BOW. Interestingly enough, we found that a NN electronic repulsion
further stabilizes this lattice/charge modulated phase. 
Under the application of an inter-chain AF coupling, we found 
new long range spin order phases showing coexisting $4k_F$ CDW or BOW
(i.e. dimerization).  
It is argued that such a simple model can well describe the various
low temperature SP, AF and SDW phases of the insulating charge 
transfer salts of the sulfur series~\cite{ESR}. 


Computations were performed at
the Supercomputer Computations Research Institute (SCRI)
and at the Academic Computing and Network Services at Tallahassee
(Florida) and at IDRIS, Orsay (France). 
Support from ECOS-SECyT A97E05 is also acknowledged.  


\begin{references}
\bibitem{note_SDW} Note that these SDW Mott-insulators
should not be confused with the weak coupling SDW phases 
of the quasi-1D (TMTSF)$_2$X metals.
\end{references}
\end{document}